\documentclass[9pt]{article}
\usepackage{spconf}
\usepackage{graphicx}
\usepackage{epstopdf}
\usepackage{bm, dsfont, algorithmic, algorithm, braket, amsmath, amssymb, cite}
\usepackage{subfig}
\usepackage{textcomp}
\usepackage{ifpdf}
\usepackage{soul}
\usepackage{url}

\begin{document}
\title{
AN APPROXIMATE MESSAGE PASSING ALGORITHM FOR RAPID PARAMETER-FREE COMPRESSED SENSING MRI
}

\name{Charles Millard$^{\star \dagger \ddagger}$, Aaron T Hess$^{\dagger}$, Boris Mailhe$^{\ddagger}$ and Jared Tanner$^{\star}$}
\address{$^{\star}$Mathematical Institute, University of Oxford,  UK \\$^{\dagger}$Oxford Centre for Clinical Magnetic Resonance, University of Oxford, UK \\ $^{\ddagger}$Siemens Healthineers, Princeton, NJ, USA \thanks{This research was supported by an EPSRC Industrial CASE studentship with Siemens Healthineers, voucher number 17000051, and through the Alan Turing Institute under the EPSRC grant EP/N510129/1. The concepts and information presented in this paper are based on research results that are not commercially available.}}

\markboth{IEEE International Conference on Image Processing 2020}
{Shell \MakeLowercase{\textit{et al.}}: Bare Demo of IEEEtran.cls for IEEE Journals}
\maketitle

\begin{abstract}
 For certain sensing matrices, the Approximate Message Passing (AMP) algorithm efficiently reconstructs undersampled signals. However, in Magnetic Resonance Imaging (MRI), where Fourier coefficients of a natural image are sampled with variable density, AMP encounters convergence problems. In response we present an algorithm based on Orthogonal AMP constructed specifically for variable density partial Fourier sensing matrices. For the first time in this setting a state evolution has been observed. A practical advantage of state evolution is that Stein's Unbiased Risk Estimate (SURE) can be effectively implemented, yielding an algorithm with no free parameters. We empirically evaluate the effectiveness of the parameter-free algorithm on simulated data and find that it converges over 5x faster and to a lower mean-squared error solution than Fast Iterative Shrinkage-Thresholding (FISTA).    
\end{abstract}


%%%%%%%%%%%%%%%%%%%%%%%%%%%%%%%%%%%%%%%%%%%
\section{Introduction}

We consider a complex linear regression problem, where complex data vector $\bm{y} \in \mathds{C}^{n}$ is formed of noisy linear measurements of a signal of interest $\bm{x}_0 \in \mathds{C}^N$:
\begin{equation}
    \bm{y} = \bm{\Phi}\bm{x}_0 + \bm{\varepsilon}, \label{eqn:SLR}
\end{equation}
where $\bm{\Phi} \in \mathds{C}^{n\times N}$ and $\bm{\varepsilon} \backsim \mathcal{CN}(\bm{0}, \sigma_\varepsilon^2 \mathds{1}_n)$, where $\mathds{1}_n$ is the $n \times n$ identity matrix. Here, $\mathcal{CN}(\bm{\mu}, \bm{\Sigma}^2)$ denotes the complex normal distribution with mean $\bm{\mu}$, covariance $\bm{\Sigma}^2$ and white phase.  A well-studied approach is to seek a solution of 
\begin{equation}
   \hat{\bm{x}} = \underset{\bm{x} \in \mathds{C}^N}{\operatorname{argmin}} \frac{1}{2} \| \bm{y} -  \bm{\Phi}\bm{x}\|^2_2 + f(\bm{x}) \label{eqn:recon}
\end{equation}
where $f(\bm{x})$ is a model-based penalty function. Compressed sensing \cite{Donoho2006CompressedSensingb, Candes2006RobustInformation} concerns the reconstruction of $\bm{x}_0$ from underdetermined measurements $n < N$. Commonly sparsity in $\hat{\bm{x}}$ is promoted by solving \eqref{eqn:recon} with $f(\bm{x}) = \lambda \|\bm{\Psi x}\|_1$ for sparse weighting $\lambda > 0$ and sparsifying transform $\bm{\Psi}$. 

The Approximate Message Passing (AMP) algorithm  \cite{Donoho2009Message-passingSensing.} is an iterative method that estimates $\bm{x}_0$ in linear regression problems such as \eqref{eqn:SLR}.  At iteration $k$ AMP implements a denoiser $\bm{g}(\bm{r}_k; \tau_k)$ on $\bm{x}_0$ estimate $\bm{r}_{k}$ with mean-squared error estimate $\tau_k$. For instance, for problems of the form of \eqref{eqn:recon}, $\bm{g}(\bm{r}_k; \tau_k)$ is the proximal operator associated with $f(\bm{x})$:
\begin{equation}
    \bm{g}(\bm{r}_k; \tau_k) = \underset{\bm{x} \in \mathds{C}^N}{\operatorname{argmin}} \frac{1}{2\tau_k}\|\bm{r}_k - \bm{x}\|^2_2 + f(\bm{x}), 
\end{equation}
which is equal to soft thresholding in the case of $f(\bm{x}) = \lambda \|\bm{\Psi x}\|_1$ and orthogonal $\bm{\Psi}$. For certain sensing matrices and given mild conditions on $f(\bm{x})$, AMP's state evolution guarantees that in the large system limit $n,N \rightarrow \infty$, $n/N \rightarrow \delta \in (0,1)$, vector $\bm{r}_{k}$  behaves as the original signal corrupted by zero-mean white Gaussian noise:  
\begin{equation}
    \bm{r}_{k} = \bm{x}_0 + \mathcal{CN}(\bm{0}, \sigma_k^2\mathds{1}_N) \label{eqn:SE}
\end{equation}
where  $\sigma_k$ is an iteration-dependant standard deviation. The state evolution of AMP has been proven for real i.i.d. Gaussian measurements in \cite{Bayati2011TheSensing} and i.i.d. sub-Gaussian measurements in \cite{Bayati2015UniversalityAlgorithms}. It has also been empirically shown that state evolution holds for uniformly undersampled Fourier measurements of a random artificial signal \cite{Donoho2009Message-passingSensing.}. When state evolution holds, AMP is known to exhibit very fast convergence. However, for generic $\bm{\Phi}$, the behavior of AMP is not well understood and it has been noted by a number of authors \cite{Rangan2014OnMatrices,Caltagirone2014OnPassingb, Rangan2016FixedMatrices} that it can encounter convergence problems. The recent Vector Approximate Message Passing (VAMP) \cite{Rangan2019} algorithm and related Orthogonal AMP (OAMP) \cite{Ma2017}  obey \eqref{eqn:SE} for a broader class of measurement matrices $\bm{\Phi}$, and were found to perform very well on certain reconstruction tasks. For VAMP, state evolution was proven for sensing matrices that are `right-orthogonally invariant': see \cite{Rangan2019} for details.

% \subsection{AMP for compressed sensing MRI}

In compressed sensing MRI \cite{Lustig2007}, measurements are formed of undersampled Fourier coefficients, so that $\bm{\Phi} = \bm{M}_\Omega \bm{F}$, where $\bm{F}$ is a 2D or 3D discrete Fourier transform and $\bm{M}_\Omega \in \mathds{R}^{n \times N}$ is a undersampling mask that selects the $j$th row of $\bm{F}$ if $j \in \Omega$ for sampling set $\Omega$. The signal of interest $\bm{x}_0$ is a natural image, so typically has a highly non-uniform spectral density that is concentrated at low frequencies. Accordingly, the sampling set $\Omega$ is usually generated such that there is a higher probability of sampling lower frequencies. This work considers an $\Omega$ with elements drawn independently from a Bernoulli distribution with non-uniform probability, such that $\mathrm{Prob}(j \in \Omega) = p_{j} \in [0,1]$. In this variable density setting there are no theoretical guarantees for AMP, VAMP and OAMP and in practice the behavior of \eqref{eqn:SE} is not observed and the algorithms typically perform poorly.  VAMP for Image Recovery (VAMPire) \cite{Schniter2017a} is an adaption of VAMP for variable density Fourier sampled images that ``whitens" the signal in the wavelet domain with a hand-tuned prediction of the wavelet energy, but like the aforementioned algorithms does not obey \eqref{eqn:SE}, as discussed in Section 2. 

Herein we present a method for undersampled signal reconstruction based on OAMP \cite{Ma2017} which we term the Variable Density Approximate Message Passing (VDAMP) algorithm. For Fourier coefficients of a realistic image randomly sampled with generic variable density and orthogonal wavelet $\bm{\Psi}$ we have empirical evidence that a state evolution occurs. Unlike the white effective noise of \eqref{eqn:SE}, the $\bm{r}_{k}$ of VDAMP behaves as the ground truth corrupted by zero-mean Gaussian noise with a separate variance for each wavelet subband, such that 
\begin{equation}
 \bm{r}_{k} = \bm{w}_0 + \mathcal{CN}(\bm{0},  \bm{\Sigma}_k^2 ), \label{eqn:VDAMPse}
\end{equation}
where $\bm{w}_0 := \bm{\Psi x}_0$ and the effective noise covariance $\bm{\Sigma}_k^2$ is diagonal so that for a $\bm{\Psi}$ with $s$ decomposition scales
\begin{equation}
    \bm{\Sigma}^2_k = \begin{bmatrix}
    \sigma^2_{k, 1} \mathds{1}_{N_{1}} & 0 & \dots & 0 \\
    0       &  \sigma^2_{k, 2} \mathds{1}_{N_2}& \dots & 0 \\
    \vdots       &  \vdots & \ddots &  \vdots \\
    0       & 0 & \hdots &  \sigma^2_{k, 1+3s}\mathds{1}_{N_{1+3s}}
    \end{bmatrix}, \label{eqn:sigma}
\end{equation}
where $\sigma^2_{k, j}$ and $N_j$ refer to the variance and dimension of the $j$th subband respectively. We refer to \eqref{eqn:VDAMPse} as the \textit{colored} state evolution of VDAMP. 

Selecting appropriate regularisation parameters such as $\lambda$ is a notable challenge in real-world compressed sensing MRI applications. We present an approach to parameter-free compressed sensing reconstruction using Stein's Unbiased Risk Estimate (SURE) \cite{Stein1981} in conjunction with VDAMP, building on the work on AMP introduced in \cite{Mousavi2013ParameterlessPassing}. A strength of automatic parameter tuning via SURE is that the it is possible to have a richer regularizer than would usually be feasible for a hand-tuned $f(\bm{x})$ \cite{Deledalle2014}. We implement a variation on the SureShrink denoiser \cite{Donoho1995AdaptingShrinkage}, using a iteration-dependant regularizer that has a separate sparse weighting per subband: 
% \begin{equation}
%  \bm{\lambda}_k = \begin{bmatrix}
%     \lambda_{k,1} \bm{1}_{N_1} \\
%     \lambda_{k,2} \bm{1}_{N_2} \\
%     \vdots \\
%     \lambda_{k,1+3s} \bm{1}_{N_{1+3s}}
%     \end{bmatrix}   \label{eqn:veclambda}
% \end{equation}
\begin{equation}
 \bm{\lambda}_k = \begin{bmatrix}
    \lambda_{k,1} \bm{1}_{N_1}^H &
    \lambda_{k,2} \bm{1}_{N_2}^H&
    \ldots &
    \lambda_{k,1+3s} \bm{1}_{N_{1+3s}}^H
    \end{bmatrix}^H,  \label{eqn:veclambda}
\end{equation}
where $\bm{1}_M$ is the $M$-dimensional column vector of ones. SURE has previously been employed for parameter-free compressed sensing MRI in  \cite{Khare2012}, where the Fast Iterative Shrinking-Thresholding Algorithm (FISTA) algorithm \cite{Beck2009} was used with SureShrink in place of the usual shrinkage step. This algorithm is herein referred to as SURE-IT. The effective noise of SURE-IT is highly non-Gaussian, so deviates from a proper theoretical basis for using SURE for threshold selection. To our knowledge, VDAMP is the first algorithm for variable density Fourier sampling of natural images where a state evolution has been observed. 

%%%%%%%%%%%%%%%%%%%%%%%%%%%%%%%%%%%%%%%%%%%%%%%%%%%%%%%%%%%%%%
\section{Description of algorithm \label{sec:VDAMP}}

 For AMP, VAMP and OAMP, \eqref{eqn:SE} states that the effective noise $\bm{r}_k - \bm{x}_0$ is white, so can be fully characterised by a scalar $\tau_k$. This is appropriate for the kind of uniform measurement matrices and separable, identical sparse signal models $f(\bm{x})$ that are often encountered in abstract compressed sensing problems. However, when Fourier coefficients of a natural image are sampled the effective noise is colored, so is poorly represented by a scalar \cite{Virtue2017TheReconstruction}. {While VAMPire attempts to whiten the effective noise, we propose allowing the effective noise to be colored, and present a method for modelling the color with a vector $\bm{\tau}_k$ that has one real number per wavelet subband.}
 
 \begin{algorithm}[t]
\caption{VDAMP \label{alg:VDAMP}}
\textbf{Require:} Sensing matrix $\bm{\Phi}$, orthogonal wavelet $\bm{\Psi}$, probability matrix $\bm{P}$, measurements $\bm{y}$, denoiser $\bm{g}(\bm{r}_{k}; \bm{\tau}_{k})$ and number of iterations $K_{it}$.
\begin{algorithmic}[1]
\STATE Set $\widetilde{\bm{r}}_{0} = \bm{0}$ and compute $\bm{S} = |\bm{\Phi}\bm{\Psi}^H|^2$
\FOR {$k =0,1,\ldots, K_{it}-1$} 
\STATE $\bm{z}_k = \bm{y} - \bm{\Phi}\bm{\Psi}^H\widetilde{\bm{r}}_{k}$ 
\STATE $\bm{r}_{k} = \widetilde{\bm{r}}_{k} + \bm{\Psi} \bm{\Phi}^H\bm{P}^{-1}\bm{z}_k$
\STATE $\bm{\tau}_{k} = \bm{S}^H \bm{P}^{-1} [(\bm{P}^{-1} - \mathds{1}_n)|\bm{z}_k|^2 + \sigma_\varepsilon^2 \bm{1}_n] $
\STATE $\hat{\bm{w}}_{k} = \bm{g}(\bm{r}_{k}; \bm{\tau}_{k})$
\STATE $\bm{\alpha}_{k} = \braket{\bm{g}'(\bm{r}_{k}; \bm{\tau}_{k})}_{\text{sband}}$
\STATE $\widetilde{\bm{r}}_{k+1} = (\hat{\bm{w}}_{k} - \bm{\alpha}_k \odot \bm{r}_{k})\oslash (\bm{1}-\bm{\alpha}_k)$
\ENDFOR
\RETURN $\hat{\bm{x}} = \bm{\Psi}^H\bm{w}_k + \bm{\Phi}^H(\bm{y} - \bm{\Phi} \bm{\Psi}^H \bm{w}_k)$
\end{algorithmic}
\end{algorithm}

The VDAMP algorithm is shown in Algorithm \ref{alg:VDAMP}. Here, $\bm{P} \in \mathds{R}^{n \times n}$ is the diagonal matrix formed of sampling probabilities $p_j$ for $j \in \Omega$. The function $\bm{g}(\bm{r}_{k}; \bm{\tau}_{k})$ refers to some denoiser with a colored effective noise model. The notation $\braket{\bm{g}'(\bm{r}_{k}; \bm{\tau}_{k})}_{\text{sband}}$ in line 7 refers to the (sub)-gradient of the denoiser averaged over subbands, so that for $s$ decomposition scales $\bm{\alpha}_{k}$ has $1+3s$ unique entries, having the same structure as the $\bm{\lambda}_k$ of  \eqref{eqn:veclambda}. In line 8, the notation $\odot$ is used for entry-wise multiplication and $\oslash$ for entry-wise division. $|\cdot|$ refers to the entry-wise absolute magnitude of a vector or matrix.

To ensure that $\bm{r}_{k}$ is an unbiased estimate of $\bm{x}_0$, the sensing matrix must be correctly normalized. In VDAMP this is manifest in the gradient step of lines 3-4, which features a crucial weighting by $\bm{P}^{-1}$. {As an intuitive example, consider $k=0$, where $\bm{r}_0 = \bm{\Psi \Phi}^H\bm{P}^{-1}\bm{y}$. Since $\mathds{E}_{\Omega, \varepsilon} (\bm{P}^{-1}\bm{y}) = \bm{y}_0$, it follows that $\bm{r}_0$ is unbiased:  $\mathds{E}(\bm{r}_0) = \bm{w}_0$.} Such a rescaling is referred to as ``density compensation" in the MRI literature \cite{Pipe1999}, and was used in the original compressed sensing MRI paper with zero-filling  to generate a unregularized, non-iterative baseline \cite{Lustig2007}. {VDAMP's density compensation contrasts with VAMPire's approach, where the sensing matrix is right-multiplied by a diagonal matrix of predicted wavelet coefficient energies. This leads to a biased $\bm{r}_k$ update, $\mathds{E}(\bm{r}_k) \neq \bm{w}_0$, and consequently it does not obey a state evolution.}

Line 5 of Algorithm \ref{alg:VDAMP} computes an estimate of the colored effective noise variance $|\bm{w}_0 - \bm{r}_k|^2$. The mean-squared error estimate $\bm{\tau}_k$ is an unbiased estimate of the expected entry-wise squared error, $
    \mathds{E}_{\Omega, \bm{\varepsilon}}(\bm{\tau}_k)  =   \mathds{E}_{\Omega, \bm{\varepsilon}}(|\bm{w}_0 - \bm{r}_k|^2),  \label{eqn:tauexp}$ 
for any $\widetilde{\bm{r}}_k$. We assume this estimator concentrates around its expectation, and leave the study of the constraints this imposes on $\bm{P}$ for future works. Note that $\bm{S}$ has $1+3s$ unique columns, so for fixed $s$ the complexity of VDAMP is governed by $\bm{\Psi}$ and $\bm{\Phi}$, whose fast implementations have complexity $O(NlogN)$. Lines 6-8 are the model-based regularization phase from OAMP and VAMP, but with a colored effective noise model.  Line 10 implements an unweighted gradient step that enforces exact consistency of the image estimate with the measurements. 

\begin{figure}[t]
\centering
    \includegraphics[width=0.95\columnwidth]{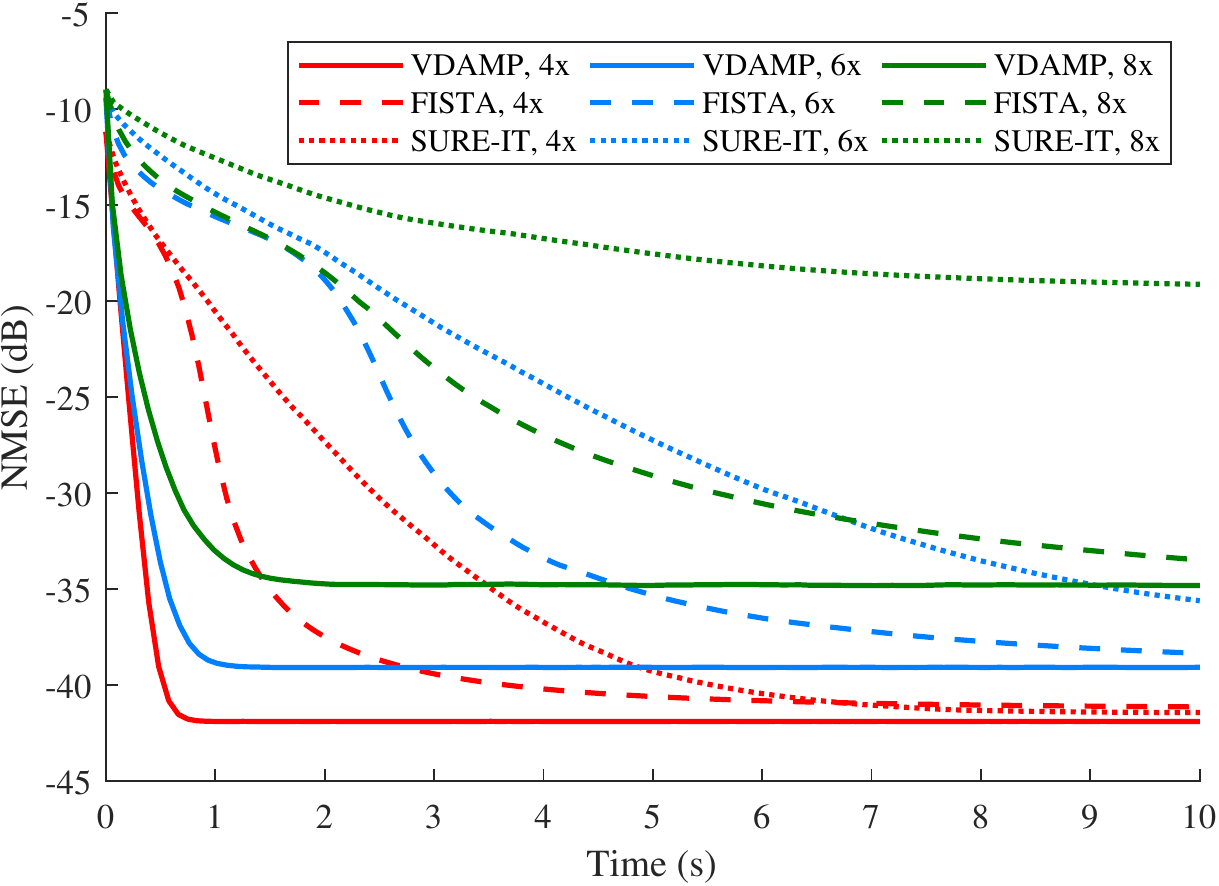}
    \caption{NMSE of VDAMP, FISTA and SURE-IT versus compute time for undersampling factors 4, 6 and 8.}
    \label{fig:VDAMPvsFISTA}
\end{figure}

%%%%%%%%%%%%%%%%%%%%%%%%%%%%%%%%%%%%%%%%%%%%%%%%%%%%%%%%%%%%%%%%%%
\section{Numerical experiments \label{sec:NumExp}}

In the experiments presented in this section the denoiser $\bm{g}(\bm{r}_{k}; \bm{\tau}_{k})$ was the complex soft thresholding operator with an automatically tuned subband-dependant threshold. In other words, we used a complex, colored version of SURE to approximately solve
\begin{equation*}
        \bm{g}(\bm{r}_{k}; \bm{\tau}_{k}) \approx \underset{ \bm{w}  \in \mathds{C}^N}{\operatorname{argmin}} \underset{\bm{\lambda}  \in \mathds{R}^N}{\operatorname{min}} \frac{1}{2} \|(\bm{w} - \bm{r}_{k})\oslash \sqrt{\bm{\tau}_{k}}\|^2_2  + \|\bm{\lambda} \odot \bm{w}\|_1, \label{eqn:gSURE}
\end{equation*}
where $\sqrt{\bm{\tau}_{k}}$ is the entry-wise square root of $\bm{\tau}_{k}$ and $\bm{\lambda}$ is of the form of \eqref{eqn:veclambda}. 

We considered the reconstruction of a $512 \times 512$ Shepp-Logan artificially corrupted with complex Gaussian noise with white phase so that $\|\bm{F x}_0\|^2_2/N\sigma_\varepsilon^2 = 40\mathrm{dB}$. We assumed that $\sigma_\varepsilon^2$ was known a priori; in practice it can be well estimated with an empty prescan. All sampling probabilities $p_j$ were generated with polynomial variable density. We considered a Haar wavelet $\bm{\Psi}$ at $s=4$ decomposition scales, which are referred to as scales 1-4, where scale 1 is the finest and scale 4 is the coarsest. All experiments were conducted in MATLAB 9.4 and can be reproduced with code available at https://github.com/charlesmillard/VDAMP.
\begin{figure}[t]
  \includegraphics[width=1.0\columnwidth]{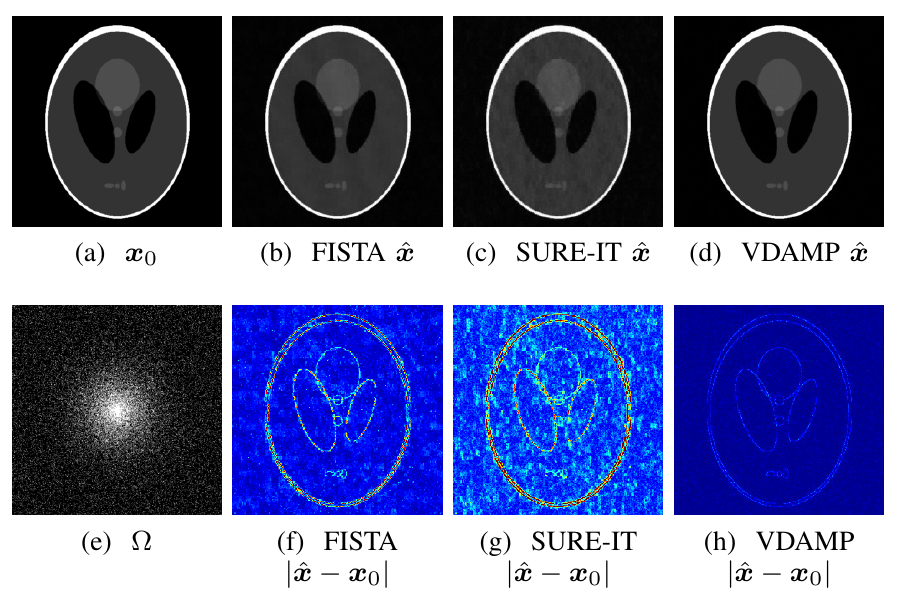}
\caption{\label{fig:x_recon} Reconstruction of a 8x undersampled Shepp-Logan after 2 seconds with (b) FISTA (NMSE = -19.3dB) (c) SURE-IT (NMSE = -16.6dB) and (d) VDAMP (NMSE -34.9dB), where absolute values are shown.} %For visualization, in (f), (g) and (h) the intensities range from 0 to 25\% of the maximum of SURE-IT's $|\hat{\bm{x}} - \bm{x}_0|$.}
\end{figure}

VDAMP was compared with FISTA, which is commonly used for compressed sensing MRI in practise, and SURE-IT, which contrasts our use of SURE. We did not compare with VAMPire as no implementation was available. For FISTA we used a sparse weighting $\lambda$ tuned with an exhaustive search so that the mean-squared error was minimised after 10 seconds. For SURE-IT the mean-squared error estimate was updated using the ground truth: $\tau_k = \|\bm{w}_0 - \bm{r}_k\|^2_2/N$, and \eqref{eqn:gSURE} with $\bm{\tau}_k = \tau_k \bm{1}_N$ was implemented.  All algorithms were initialised with a vector of zeros. Three sampling sets $\Omega$ were generated at undersampling factors of approximately 4, 6 and 8, and VDAMP, FISTA and SURE-IT were run for 10 seconds.

 Fig. \ref{fig:VDAMPvsFISTA} shows the NMSE $= \|\hat{\bm{x}}- \bm{x}_0\|^2_2/\|\bm{x}_0\|^2_2$ as a function of time for each algorithm.  The mean per-iteration compute time was 0.065s for FISTA, 0.077s for SURE-IT, and 0.091s for VDAMP. Fig. \ref{fig:x_recon} shows the ground truth image, sampling set, and FISTA and VDAMP reconstruction at undersampling factor 8 after 2s, with entry-wise errors $|\hat{\bm{x}} - \bm{x}_0|$.
\begin{figure}
% \captionsetup[subfigure]{labelformat=empty}

    \centering \includegraphics[width=1\columnwidth]{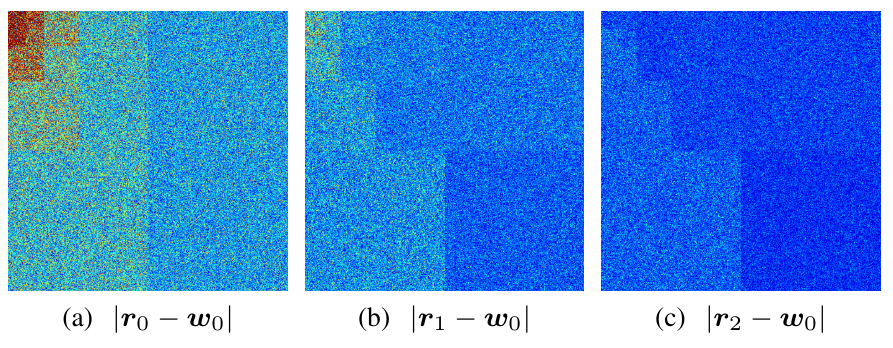} 
\caption{\label{fig:wav_err} $|\bm{r}_{k} - \bm{w}_0|$ of VDAMP for $k=0$, $k= 1$ and $k=2$.}
\vspace{1cm}
    \includegraphics[width=1\columnwidth]{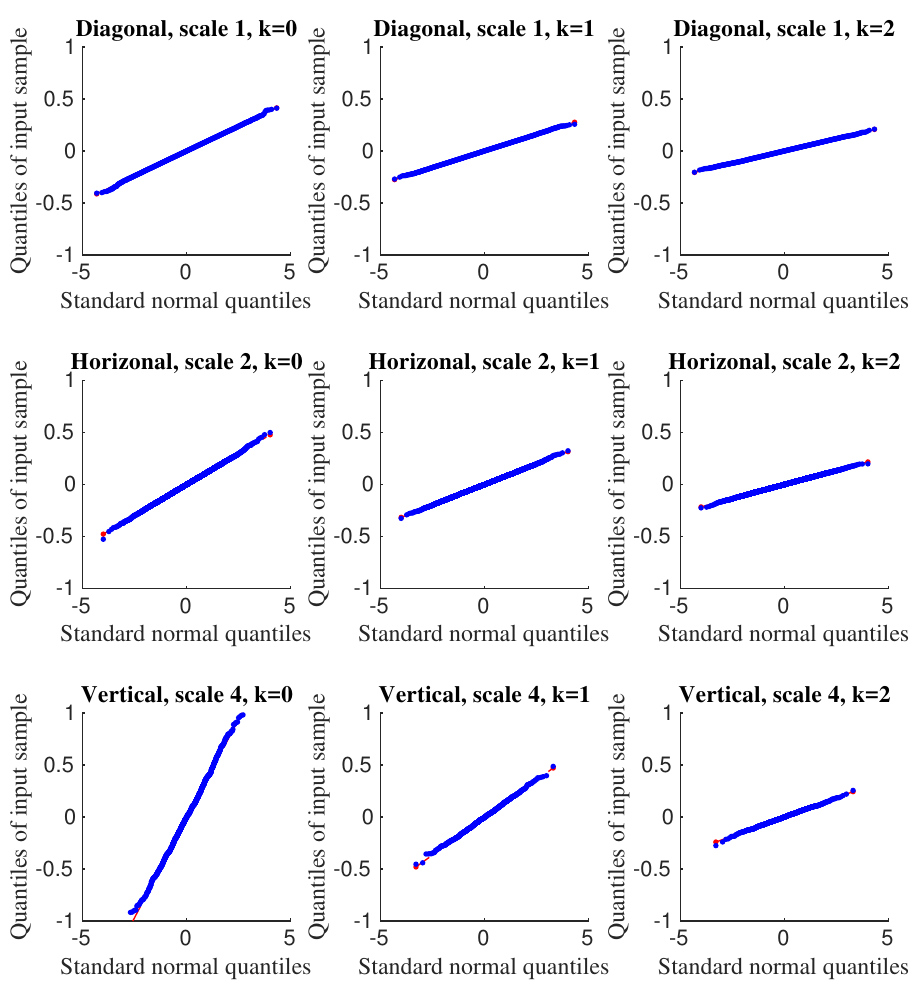}
    \centering
    \caption{Normalized quantile-quantile plots against a Gaussian of three subbands of $\bm{r}_{k}-\bm{w}_0$ for $k=0$, $k=1$ and $k=2$ in blue, and points along a straight line in red. The real part is plotted in the top and bottom rows and the imaginary is plotted in the middle row. Linearity of the blue points indicates that that the data comes from a Gaussian distribution.  Finite dimensional effects causing small deviations from an exact Gaussian are more apparent at coarse scales, where the dimension is smaller.}
    \label{fig:qqplots}
\end{figure}
In Fig. \ref{fig:wav_err}, the absolute value of the residual of $\bm{r}_{k}$ is shown for three representative iterations: $k=0$, $k=1$ and $k=2$ for undersampling factor 8.  For the same iterations, Fig. \ref{fig:qqplots} shows quantile-quantile plots of the real parts of three illustrative subbands of $\bm{r}_{k} - \bm{w}_0$: the diagonal detail at scale 1, the horizontal detail at scale 2 and the vertical detail at scale 4. These figures provide empirical evidence that the effective noise of VDAMP evolves as \eqref{eqn:VDAMPse}. 

\begin{figure}
\centering
    \includegraphics[width=0.97\columnwidth]{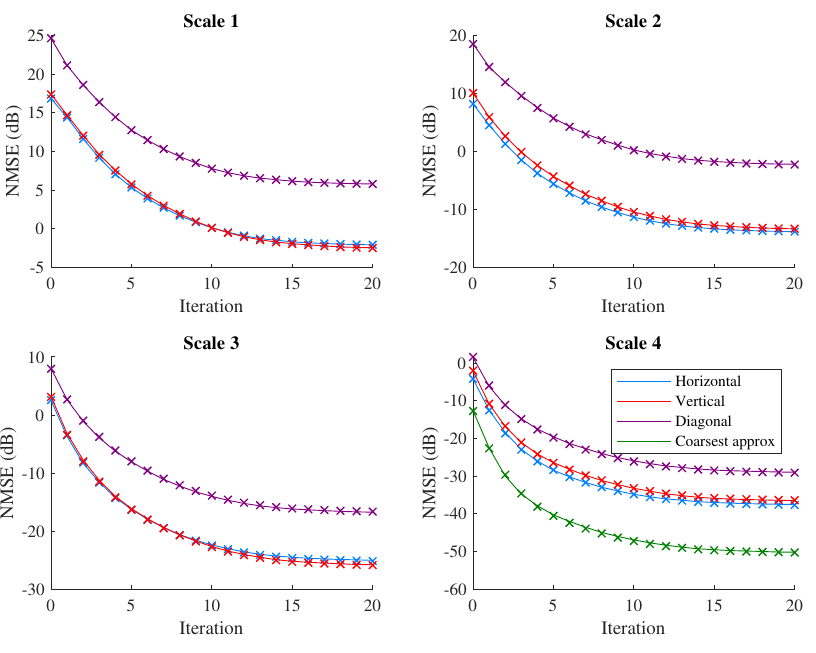}
    \caption{NMSE versus iteration number of $\bm{r}_{k}$ for each subband. Lines show the actual NMSE and crosses show the predictions from $\bm{\tau}_{k}$.}
    \label{fig:bwerr}
\end{figure}

The efficacy of automatic threshold selection with SURE depends on how accurately the diagonal of $\bm{\Sigma}_k^2$ from \eqref{eqn:sigma} is modelled by $\bm{\tau}_k$. For $k = 0, 1, \ldots, 20,$ Fig. \ref{fig:bwerr} shows the ground truth NMSE of the wavelet coefficients at all four scales and the prediction of VDAMP, where the NMSE is per subband. %The accuracy of VDAMP's NMSE estimation ensures that for sufficiently large $N_v$ the threshold selection $\hat{t}$ of \eqref{eqn:that} is near-optimal. 

\section{Conclusions }
 VDAMP's state evolution provides an informed and efficient way to tune model parameters via SURE.  More degrees of freedom are feasibly allowed in the model, enabling higher order prior information such as anisotropy, variability across scales and structured sparsity, without the need to estimate the structure a priori such as in model-based compressed sensing \cite{Baraniuk2010Model-BasedSensing}.   Theoretical work is required to establish the generality of the state evolution empirically observed in these experiments.

It is known that the state evolution of AMP holds for a wide range of denoisers $\bm{g}(\bm{r}_{k}; \tau_{k})$ \cite{Metzler2016FromSensing}. In a recent survey \cite{Ahmad2019PlugImaging} a number of standard compressed sensing algorithms that leverage image denoisers designed for Gaussian noise were shown to perform well on MRI reconstruction tasks, despite the mismatch between the effective noise and its model. A sophisticated denoiser equipped to deal with wavelet coefficients corrupted with known colored Gaussian noise would be expected to perform well in conjunction with VDAMP.

\clearpage

\bibliographystyle{IEEE}
\bibliography{main}

\begin{thebibliography}{10}

\bibitem{Donoho2006CompressedSensingb}
David~L Donoho,
\newblock ``{Compressed sensing},''
\newblock {\em IEEE Transactions on Information Theory}, vol. 52, no. 4, pp.
  1289--1306, 4 2006.

\bibitem{Candes2006RobustInformation}
Emmanuel Candes, Justin Romberg, and Terence Tao,
\newblock ``{Robust uncertainty principles: exact signal reconstruction from
  highly incomplete frequency information},''
\newblock {\em IEEE Transactions on Information Theory}, vol. 52, no. 2, pp.
  489--509, 2 2006.

\bibitem{Donoho2009Message-passingSensing.}
David~L Donoho, Arian Maleki, and Andrea Montanari,
\newblock ``{Message-passing algorithms for compressed sensing.},''
\newblock {\em Proceedings of the National Academy of Sciences of the United
  States of America}, vol. 106, no. 45, pp. 18914--9, 11 2009.

\bibitem{Bayati2011TheSensing}
Mohsen Bayati and Andrea Montanari,
\newblock ``{The Dynamics of Message Passing on Dense Graphs, with Applications
  to Compressed Sensing},''
\newblock {\em IEEE Transactions on Information Theory}, vol. 57, no. 2, pp.
  764--785, 2 2011.

\bibitem{Bayati2015UniversalityAlgorithms}
Mohsen Bayati, Marc Lelarge, and Andrea Montanari,
\newblock ``{Universality in polytope phase transitions and message passing
  algorithms},''
\newblock {\em The Annals of Applied Probability}, vol. 25, no. 2, pp.
  753--822, 4 2015.

\bibitem{Rangan2014OnMatrices}
Sundeep Rangan, Philip Schniter, and Alyson Fletcher,
\newblock ``{On the convergence of approximate message passing with arbitrary
  matrices},''
\newblock in {\em 2014 IEEE International Symposium on Information Theory}. 6
  2014, pp. 236--240, IEEE.

\bibitem{Caltagirone2014OnPassingb}
Francesco Caltagirone, Lenka Zdeborova, and Florent Krzakala,
\newblock ``{On convergence of approximate message passing},''
\newblock in {\em 2014 IEEE International Symposium on Information Theory}. 6
  2014, pp. 1812--1816, IEEE.

\bibitem{Rangan2016FixedMatrices}
Sundeep Rangan, Philip Schniter, Erwin Riegler, Alyson~K. Fletcher, and Volkan
  Cevher,
\newblock ``{Fixed Points of Generalized Approximate Message Passing With
  Arbitrary Matrices},''
\newblock {\em IEEE Transactions on Information Theory}, vol. 62, no. 12, pp.
  7464--7474, 12 2016.

\bibitem{Rangan2019}
Sundeep Rangan, Philip Schniter, and Alyson~K. Fletcher,
\newblock ``{Vector Approximate Message Passing},''
\newblock {\em IEEE Transactions on Information Theory}, pp. 1--1, 2019.

\bibitem{Ma2017}
Junjie Ma and Li~Ping,
\newblock ``{Orthogonal AMP},''
\newblock {\em IEEE Access}, vol. 5, pp. 2020--2033, 2017.

\bibitem{Lustig2007}
Michael Lustig, David Donoho, and John~M. Pauly,
\newblock ``{Sparse MRI: The application of compressed sensing for rapid MR
  imaging},''
\newblock {\em Magnetic Resonance in Medicine}, vol. 58, no. 6, pp. 1182--1195,
  12 2007.

\bibitem{Schniter2017a}
Philip Schniter, Sundeep Rangan, and Alyson Fletcher,
\newblock ``{Plug-and-play Image Recovery using Vector AMP},''
\newblock 2017, BASP Frontiers Workshop 2017.

\bibitem{Stein1981}
Charles~M. Stein,
\newblock ``{Estimation of the Mean of a Multivariate Normal Distribution},''
\newblock {\em The Annals of Statistics}, vol. 9, no. 6, pp. 1135--1151, 11
  1981.

\bibitem{Mousavi2013ParameterlessPassing}
Ali Mousavi, Arian Maleki, and Richard~G. Baraniuk,
\newblock ``{Parameterless Optimal Approximate Message Passing},''
\newblock 10 2013.

\bibitem{Deledalle2014}
Charles~Alban Deledalle, Samuel Vaiter, Jalal Fadili, and Gabriel Peyr{\'{e}},
\newblock ``{Stein Unbiased GrAdient estimator of the Risk (SUGAR) for multiple
  parameter selection},''
\newblock {\em SIAM Journal on Imaging Sciences}, vol. 7, no. 4, pp.
  2448--2487, 2014.

\bibitem{Donoho1995AdaptingShrinkage}
David~L. Donoho and Iain~M. Johnstone,
\newblock ``{Adapting to Unknown Smoothness via Wavelet Shrinkage},''
\newblock {\em Journal of the American Statistical Association}, vol. 90, no.
  432, pp. 1200--1224, 12 1995.

\bibitem{Khare2012}
Kedar Khare, Christopher~J. Hardy, Kevin~F. King, Patrick~A. Turski, and Luca
  Marinelli,
\newblock ``{Accelerated MR imaging using compressive sensing with no free
  parameters},''
\newblock {\em Magnetic Resonance in Medicine}, vol. 68, no. 5, pp. 1450--1457,
  11 2012.

\bibitem{Beck2009}
Amir Beck and Marc Teboulle,
\newblock ``{A Fast Iterative Shrinkage-Thresholding Algorithm for Linear
  Inverse Problems},''
\newblock {\em SIAM Journal on Imaging Sciences}, vol. 2, no. 1, pp. 183--202,
  1 2009.

\bibitem{Virtue2017TheReconstruction}
Patrick Virtue and Michael Lustig,
\newblock ``{The Empirical Effect of Gaussian Noise in Undersampled MRI
  Reconstruction},''
\newblock {\em Tomography}, vol. 3, no. 4, pp. 211--221, 12 2017.

\bibitem{Pipe1999}
James~G. Pipe and Padmanabhan Menon,
\newblock ``{Sampling density compensation in MRI: Rationale and an iterative
  numerical solution},''
\newblock {\em Magnetic Resonance in Medicine}, vol. 41, no. 1, pp. 179--186, 1
  1999.

\bibitem{Baraniuk2010Model-BasedSensing}
Richard~G. Baraniuk, Volkan Cevher, Marco~F. Duarte, and Chinmay Hegde,
\newblock ``{Model-Based Compressive Sensing},''
\newblock {\em IEEE Transactions on Information Theory}, vol. 56, no. 4, pp.
  1982--2001, 4 2010.

\bibitem{Metzler2016FromSensing}
Christopher~A. Metzler, Arian Maleki, and Richard~G. Baraniuk,
\newblock ``{From Denoising to Compressed Sensing},''
\newblock {\em IEEE Transactions on Information Theory}, vol. 62, no. 9, pp.
  5117--5144, 9 2016.

\bibitem{Ahmad2019PlugImaging}
Rizwan Ahmad, Charles~A. Bouman, Gregery~T. Buzzard, Stanley Chan, Edward~T.
  Reehorst, and Philip Schniter,
\newblock ``{Plug and play methods for magnetic resonance imaging},''
\newblock 3 2019.

\end{thebibliography}

\end{document}